\documentclass[pdflatex,sn-mathphys-num]{sn-jnl}

\usepackage{soul}
\usepackage{caption}
\usepackage{graphicx}%
\usepackage{multirow}%
\usepackage{amsmath,amssymb,amsfonts}%
\usepackage{amsthm}%
\usepackage{mathrsfs}%
\usepackage[title]{appendix}%
\usepackage{textcomp}%
\usepackage{manyfoot}%
\usepackage{booktabs}%
\usepackage{algorithm}%
\usepackage{algorithmicx}%
\usepackage{algpseudocode}%
\usepackage{listings}%
\usepackage{upgreek}
\usepackage{indentfirst} %
\usepackage{color}
\usepackage{xcolor}%
\usepackage{bm}

\soulregister\cite7
\theoremstyle{thmstyleone}%
\theoremstyle{thmstyletwo}%

\theoremstyle{thmstylethree}%

\begin{document}

\title{Anomalous optical valley  Hall dynamics of exciton-polaritions}

\author[1]{\fnm{Xingzhou} \sur{Chen}}
\equalcont{These authors contributed equally to this work.}
\author[1]{\fnm{Yuanjun} \sur{Guan}}
\equalcont{These authors contributed equally to this work.}
\author[2]{\fnm{Areg} \sur{Ghazaryan}}

\author[3]{\fnm{Shiran} \sur{Sun}}

\author[3]{\fnm{Lingxiao} \sur{Yu}}

\author[3]{\fnm{Ruitao} \sur{Lv}}

\author*[4]{\fnm{Artem} \sur{Volosniev}}\email{artem@phys.au.dk}

\author*[1,5,6]{\fnm{Zheng} \sur{Sun}}\email{zsun@lps.ecnu.edu.cn}

\author*[1,6,7]{\fnm{Jian} \sur{Wu}}\email{jwu@phy.ecnu.edu.cn}

 \affil[1]{\orgdiv{State Key Laboratory of Precision Spectroscopy}, \orgname{East China Normal University}, \orgaddress{\city{Shanghai}, \postcode{200241}, \country{China}}}

\affil[2]{\orgname{Institute of Science and Technology Austria}, \orgaddress{\street{Am Campus 1}, \city{Klosterneuburg}, \postcode{3400},  \country{Austria}}}

\affil[3]{\orgdiv{State Key Laboratory of New Ceramic Materials, Key Laboratory of Advanced Materials (MOE), School of Materials Science and Engineering}, \orgname{Tsinghua University}, \orgaddress{\city{Beijing}, \postcode{100084}, \country{China}}}

\affil[4]{\orgdiv{Department of Physics and Astronomy}, \orgname{Aarhus University}, \orgaddress{\city{Aarhus}, \postcode{DK-8000}, \country{Denmark}}}

\affil[5]{\orgdiv{Shanghai Key Laboratory of Magnetic Resonance, School of Physics and Electronic Science}, \orgname{East China Normal University}, \orgaddress{\city{Shanghai}, \postcode{200241}, \country{China}}}

\affil[6]{\orgdiv{Collaborative Innovation Center of Extreme Optics}, \orgname{Shanxi University}, \orgaddress{\city{Taiyuan}, \postcode{030006}, \state{Shangxi} \country{China}}}

\affil[7]{\orgdiv{Chongqing Key Laboratory of Precision Optics}, \orgname{Chongqing Institute of East China Normal University}, \orgaddress{\city{Chongqing}, \postcode{401121}, \country{China}}}

\abstract{The valley degree of freedom in atomically thin transition-metal dichalcogenides provides a natural binary index for information processing. Exciton–polaritons formed under strong light–matter coupling offer a promising route to overcome the limited lifetime and transport of bare valley excitons. Here we report an anomalous optical valley Hall effect in a monolayer WS${}_2$ exciton–polariton system. Using polarization- and time-resolved real-space imaging, we directly visualize a symmetry-breaking spatial separation of polaritons from opposite valleys under linearly polarized excitation, accompanied by an ultrafast Hall drift velocity on the order of $10^{5}~\mathrm{m/s}$. This behaviour cannot be accounted for by conventional cavity-induced mechanisms and instead points to a strain-induced synthetic pseudomagnetic field acting on the excitonic component of polaritons. Our results establish exciton–polaritons as a high-speed and optically accessible platform for valley transport, opening pathways towards tunable valleytronic and topological photonic devices.
}

\keywords{Anormalous Valley Hall effect, Exciton-polariton, Valleytronics}

\maketitle

\section{Introduction}

Atomically thin transition-metal dichalcogenides (TMDs) host a valley degree of freedom that is selectively addressable by circularly polarized light, providing a natural binary index for information processing and establishing a versatile platform for valleytronics~\cite{mak2012control,cao2012valley,zeng2012valley,mak2010atomically,splendiani2010emerging}. Within this framework, the valley Hall effect enables the encoding and manipulation of binary information through the valley pseudospin, offering a promising route towards low-energy information processing, valley-based logic and non-volatile memory architectures~\cite{xiao2007Valleycontrasting,schaibley2016Valleytronics,mak2018Light,vitale2018Valleytronics,unuchek2019Valleypolarized}. A central challenge, however, lies in achieving valley Hall transport at sufficiently high speeds. In conventional electronic and excitonic platforms, the large effective mass of carriers and the reliance on electrical control fundamentally limit both the speed and coherence of valley-selective transport, impeding direct access to ultrafast valley dynamics~\cite{barachati2018interacting,kulig2018exciton}. These constraints motivate the search for alternative carrier systems beyond electrons and excitons.

Exciton-polaritons -- hybrid light-matter quasiparticles formed under strong coupling between excitons and confined photonic modes -- offer a compelling route beyond these limitations~\cite{deng2010exciton,liu2015strong,dufferwiel2015exciton,su2020observation}. By combining an ultralight effective mass and long-range coherence inherited from their photonic component with strong interactions and valley selectivity from their excitonic part~\cite{sun2017direct,zhao2023exciton}, polaritons enable enhanced transport and flexible dispersion engineering. Importantly, the radiative decay channel of polaritons suppresses intervalley scattering, allowing robust valley polarization to persist up to room temperatures~\cite{dufferwiel2017valley,chen2017valley,sun2017optical}. These attributes position polaritonic platforms as promising candidates for scalable valley-optoelectronic devices~\cite{lamountain2021valley,cilibrizzi2023self}.

In polaritonic systems, previously reported optical valley Hall effects have predominantly originated from cavity-induced effective fields, such as transverse-electric–transverse-magnetic (TE–TM) splitting~\cite{lundt2019optical,kavokin2005optical}. This mechanism leads to a $\pi$-periodic Hall response that is intrinsically locked to the valley pseudospin texture, constraining both the symmetry and the tunability of valley transport. By contrast, a valley Hall response characterized by two macroscopic, spatially separated valley domains -- corresponding to a $2\pi$-periodic evolution -- would constitute a direct real-space analogue of the conventional charge and spin Hall effects. Such a binary spatial partitioning enables intuitive valley addressing and readout, and, critically, provides a well-defined experimental pathway to directly quantify the valley Hall drift velocity, a key figure of merit governing the speed of valleytronic devices.

Here we experimentally demonstrate an anomalous optical valley Hall effect in a monolayer WS${}_2$ exciton-polariton system. Polarization-resolved real-space imaging under linearly polarized excitation directly reveals a symmetry-breaking spatial separation between opposite valley populations. Owing to the ultralight effective mass of polaritons, this anomalous transport gives rise to an exceptionally high valley Hall drift velocity. The observed behaviour cannot be explained by conventional cavity-induced mechanisms and instead points to the emergence of a strain-induced synthetic pseudomagnetic field acting on the excitonic component of polaritons. Complementary measurements under circularly polarized excitation further confirm valley-dependent directional drift and an extended valley polarization lifetime. Together, our results establish exciton-polaritons as a dynamic, high-speed platform for valley Hall transport, opening new opportunities for tunable valleytronic and topological photonic devices.

\section{Results}

\begin{figure}[t!]
\centering
\includegraphics[width=\textwidth]{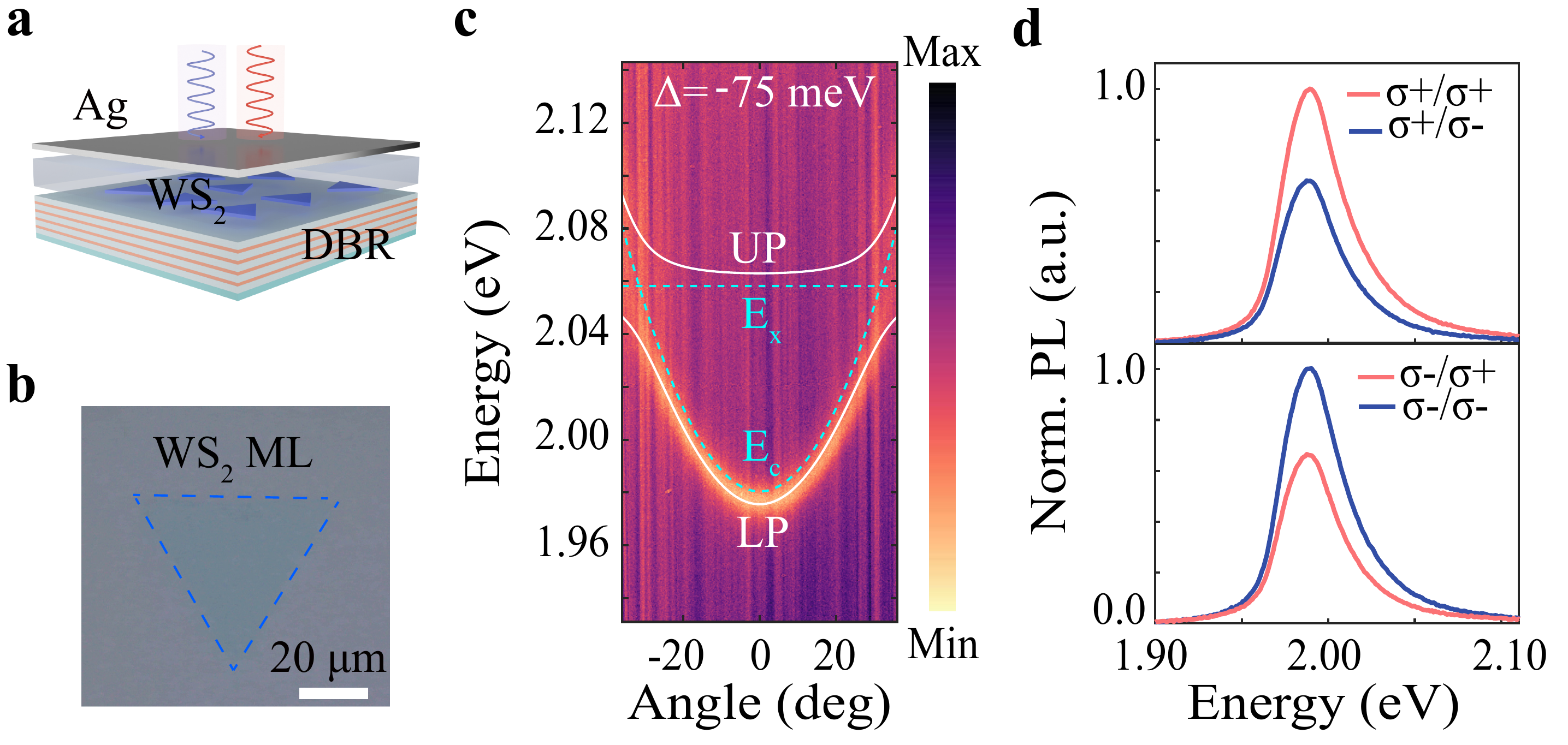}
\caption{\textbf{Sample structure and optical characterization.} \textbf{a,} Schematic of the hybrid optical microcavity consisting of a 30-pair DBR as the bottom mirror and a silver top mirror, with a monolayer WS${}_2$ embedded at the antinode of the cavity field. \textbf{b,} Optical microscope image of the cavity region. The dotted outline delineates the CVD-grown WS${}_2$ monolayer. \textbf{c,} Angle-resolved reflectance spectra of the microcavity. The solid white curves denote the upper (UP) and lower polariton (LP) branches obtained from a coupled-oscillator fit, while the blue dashed lines mark the exciton resonance, $E_X$, and the bare cavity photon dispersion, $E_c$. The extracted detuning is $\Delta \approx–75~\mathrm{meV}$. \textbf{d,} Polarization-resolved PL spectra of the lower polariton branch under $150~\mathrm{fs}$ circularly polarized $532~\mathrm{nm}$ excitation, yielding a degree of polarization of approximately $\pm25\%$.}\label{fig1}
\end{figure}

The schematic layout of the microcavity sample is shown in Fig.~\ref{fig1}a. The structure comprises a bottom distributed Bragg reflector (DBR) of 30 alternating SiO${}_2$/TiO${}_2$ pairs, followed by a $105~\mathrm{nm}$ SiO${}_2$ layer deposited by plasma-enhanced chemical vapor deposition (PECVD) to precisely tune the cavity resonance to the WS${}_2$ exciton. A CVD-grown WS${}_2$ monolayer was subsequently transferred onto the SiO${}_2$ surface~\cite{yu2024high}. To complete the cavity and protect the monolayer, a $75~\mathrm{nm}$ polymethyl methacrylate (PMMA) spacer was spin-coated and capped with a thermally evaporated $50~\mathrm{nm}$ Ag film serving as the top mirror. 

An optical micrograph of the device (Fig.~\ref{fig1}b) reveals a WS${}_2$ monolayer with lateral dimensions exceeding $60~\mathrm{\upmu m}$, providing a sufficiently large and uniform area for spatially resolved polariton measurements. Angle-resolved reflectance spectra measured at $5~\mathrm{K}$ (Fig.~\ref{fig1}c) display a clear anticrossing between the cavity photon mode and the WS${}_2$ exciton resonance, confirming operation in the strong-coupling regime. The measured dispersion is well reproduced by a coupled-oscillator model, yielding a Rabi splitting of approximately $40~\mathrm{meV}$ and a cavity–exciton detuning of $-75~\mathrm{meV}$. Although the limited numerical aperture under cryogenic conditions reduces the apparent contrast of the anticrossing, complementary room-temperature measurements (Extended Data Fig.~\ref{extended1}) reveal a more symmetric dispersion as the exciton energy shifts toward the cavity mode. Importantly, the relatively large negative detuning employed here places the lower polariton branch deep in the photon-like regime, substantially reducing its effective mass and enhancing its propagation length. This regime is particularly favourable for investigating valley-dependent transport phenomena, as it broadens the accessible energy and momentum window while enabling efficient polariton drift in real space. These characteristics form the experimental foundation for the observation of anomalous valley Hall dynamics discussed below.

To assess the valley polarization of exciton–polaritons, the microcavity was excited using a $150~\mathrm{fs}$ circularly polarized laser pulse at $532~\mathrm{nm}$. The emitted photoluminescence (PL) was analysed in both co-polarized ($\sigma^{+}/\sigma^{+}$ or $\sigma^{-}/\sigma^{-}$) and cross-polarized ($\sigma^{+}/\sigma^{-}$ or $\sigma^{-}/\sigma^{+}$) detection configurations. As shown in Fig.~\ref{fig1}d, the co-polarized PL signal dominates over the cross-polarized component, demonstrating the preservation of valley information during polariton formation and relaxation. From these measurements, we extract a degree of polarization (DOP) $\uprho = (I_{+}-I_{-})/(I_{+}+I_{-})$ of approximately $\pm25\%$, where $I_{+}$ and $I_{-}$ denote the $\sigma^{+}$ and $\sigma^{-}$ PL intensities, respectively. This value substantially exceeds that typically observed for bare excitons under comparable non-resonant excitation conditions~\cite{mak2012control,zhu2014anomalously}, indicating enhanced valley retention in the polariton regime. Such stabilization can be attributed to the hybrid light–matter character of polaritons: cavity-modified dispersion and relaxation pathways reduce intervalley scattering~\cite{dufferwiel2017valley}, while the rapid radiative decay further limits valley depolarization. These features are consistent with previous observations of cavity-protected valley pseudospins in TMD polariton systems~\cite{chen2017valley,sun2017optical}. 

\begin{figure}[t!]
\centering
\includegraphics[width=\textwidth]{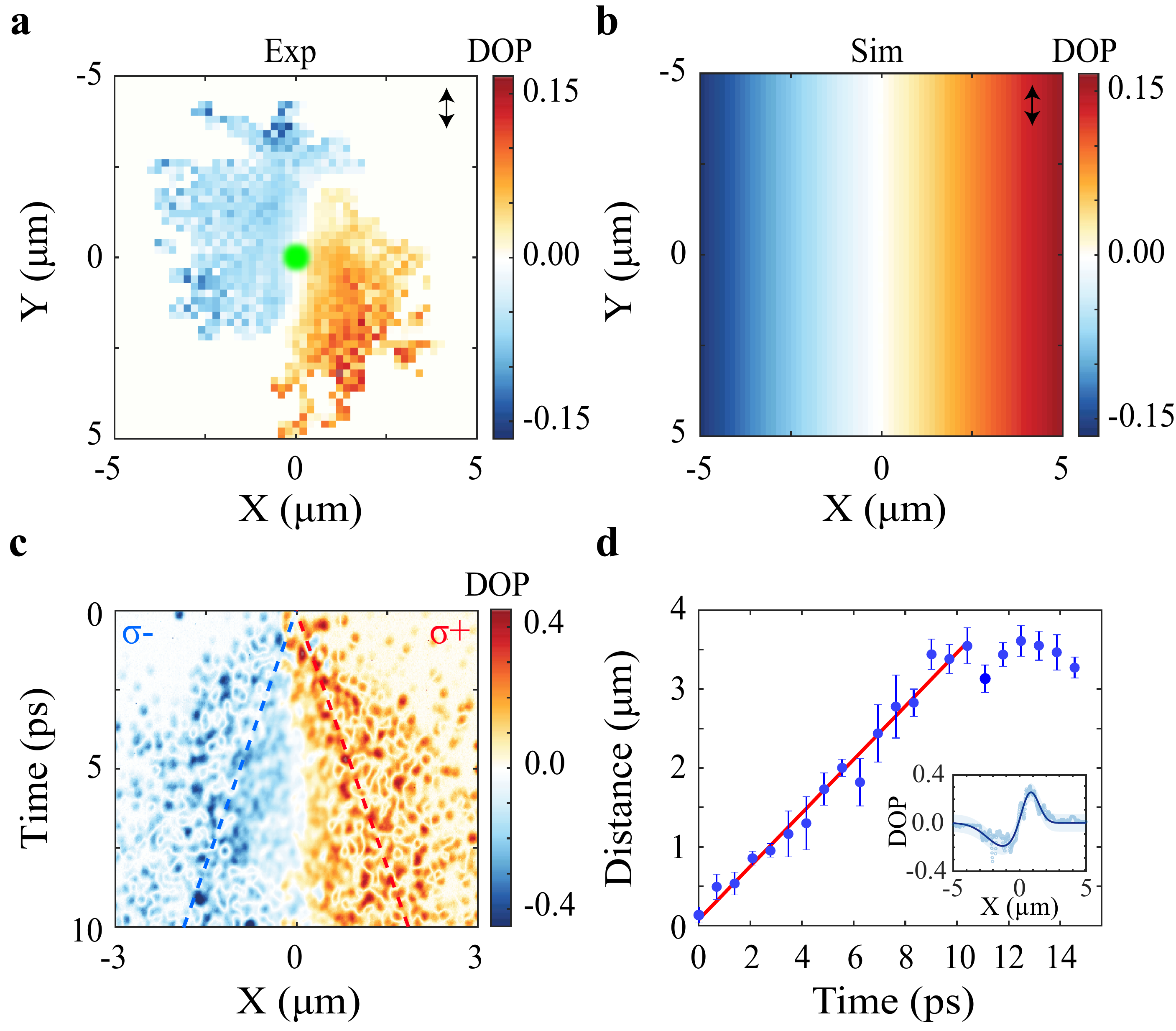}
\caption{\textbf{Dynamics of the anomalous optical valley Hall effect.} \textbf{a,} Experimental real-space map of the valley polarization under linearly (vertically) polarized excitation along the Y-direction. The green circle indicates the excitation laser spot, and the double-headed arrow denotes the polarization direction of the incident light. \textbf{b,} Simulated real-space valley-polarization map under comparable excitation conditions, incorporating strain-induced symmetry breaking via an effective pseudomagnetic field. \textbf{c,} Time-resolved real-space snapshots illustrating the evolution of the valley Hall deflection. Dashed red and blue lines indicate the evolution paths of the $\sigma^{+}$ and $\sigma^{-}$ components of the valley polarization. \textbf{d,} Displacement of the valley-polarized polariton population as a function of time at an excitation power of $30~\mathrm{mW}$. A linear fit (red line) yields a Hall drift velocity of $1.69~\times 10^{5}~\mathrm{m/s}$. The inset shows a line cut of the real-space DOP distribution at $5~\mathrm{ps}$, where the experimental data are represented by circles, the bold light-blue band indicates the $95\%$ prediction interval, and the navy-blue curve corresponds to a two-Gaussian fit.}\label{fig2}
\end{figure}

To access the optical valley Hall response, both inequivalent valleys must be populated simultaneously, which we achieve using linearly polarized excitation. The spatial distribution of valley polarization is mapped by CCD-based real-space imaging, in which $\sigma^{+}$- and $\sigma^{-}$-resolved PL are collected and combined to extract the local degree of polarization across the field of view. Figure~\ref{fig2}a shows the resulting real-space valley-polarization map under linearly polarized excitation. A pronounced spatial separation of positive and negative polarization emerges, forming a Tai Chi–like pattern. Unlike, the optical valley Hall effect previously observed in TMD polariton systems~\cite{lundt2019optical}, the polarization texture observed here lacks the $\pi$-periodic symmetry that is a defining feature of the conventional cavity-driven valley Hall effect. 

This clear symmetry-breaking points to an anomalous origin of the observed Tai Chi-like pattern, beyond photonic TE–TM splitting. Consistent with theoretical frameworks describing strain-induced valley transport, we ascribe this anomalous response to residual strain in the large-area CVD-grown WS${}_2$ monolayer, which can arise during the transfer process~\cite{yu2024high}. Such strain generates a synthetic effective magnetic field that couples to the excitonic component of the polariton wavefunction~\cite{glazov2022exciton,yagodkin2024excitons}. In this picture, the effective field components are given by $\Omega_x\sim (u_{xx}-u_{yy})$, $\Omega_y\sim u_{xy}$ and  $\Omega_z\sim (u_{xx}-u_{yy})k_x-2u_{xy}k_y$, where $u_{xx}-u_{yy}$ and $u_{xy}$ are the standard components of the strain tensor and $k_i$ is the momentum. This field drives a valley-dependent precession of the polariton population, producing an anomalous transverse deflection consistent with the observed polarization texture, see Fig.~\ref{fig2}b, where we used normal and shear strains, $u_{xx}-u_{yy}$=0.1\% and $u_{xy}=0.2$\%. To further support this interpretation, we performed measurements under different orientations of linear excitation polarization, yielding consistent symmetry-breaking responses (See Extended Data Fig.~\ref{extended5}). Details of the theoretical model are provided in the Supplementary Information.

The Hall drift velocity $v_{\text{H}}$ quantifies the transverse transport speed of carriers under a Hall response and constitutes a fundamental figure of merit that sets the intrinsic response time and operational bandwidth of Hall-based devices, including sensors, switches and topological circuit elements. Leveraging an all-optical approach, we directly track the ultrafast dynamics of valley Hall drift in real space and time. Time- and spatially resolved measurements of the $\sigma$+ and $\sigma$- PL components under linearly polarized excitation reconstruct the spatiotemporal evolution of DOP, as shown in Fig.~\ref{fig2}c. At $t=0$, the DOP is nearly uniform and close to zero, as expected for linear excitation. As time progresses, regions of positive and negative polarization clearly separate. By fitting each time frame with double Gaussians, we extract the separation distance as a function of time (Fig.~\ref{fig2}d). The separation increases linearly at early times before saturating due to the finite polariton lifetime, which limits the maximum propagation distance. A linear fit to the initial regime yields a velocity of $3.38~\times 10^{5}~\mathrm{m/s}$; half of this value defines the Hall drift velocity, $v_{\text{H}}=1.69~\times 10^{5}~\mathrm{m/s}$. This behavior is consistent with a simple ballistic model, in which $v_{\text{H}}\sim \hbar k_{\mathrm{av}}/m_p$ (Supplementary Material), where $\hbar k_{\mathrm{av}}$ is the standard deviation of the initial polariton momentum distribution, and $m_p\sim 10^{-4} m_e$ is the polariton effective mass. Using the experimental drift velocity, we estimate $ k_{\mathrm{av}}\sim 0.1~\mathrm{\upmu m^{-1}}$, in order-of-magnitude agreement with expectations for the initial polariton population. Notably, the measured Hall drift velocity exceeds typical values reported in conventional valley and spin Hall systems by one to two orders of magnitude~\cite{mak2014valley,kato2004observation}, highlighting the exceptional dynamical response enabled by the polaritonic platform. This pronounced enhancement underscores the potential of exciton–polaritons for accessing ultrafast valley transport regimes that are inaccessible in purely electronic or excitonic systems. The observed linear separation of valley-polarized populations indicates that the anomalous optical valley Hall effect establishes itself within the first picosecond, below the temporal resolution of our current measurements. Future sub-picosecond pump–probe experiments could directly capture the initial build-up of valley polarization and provide insight into the microscopic scattering processes governing the ultrafast Hall response.

\begin{figure}[t!]
\centering
\includegraphics[width=\textwidth]{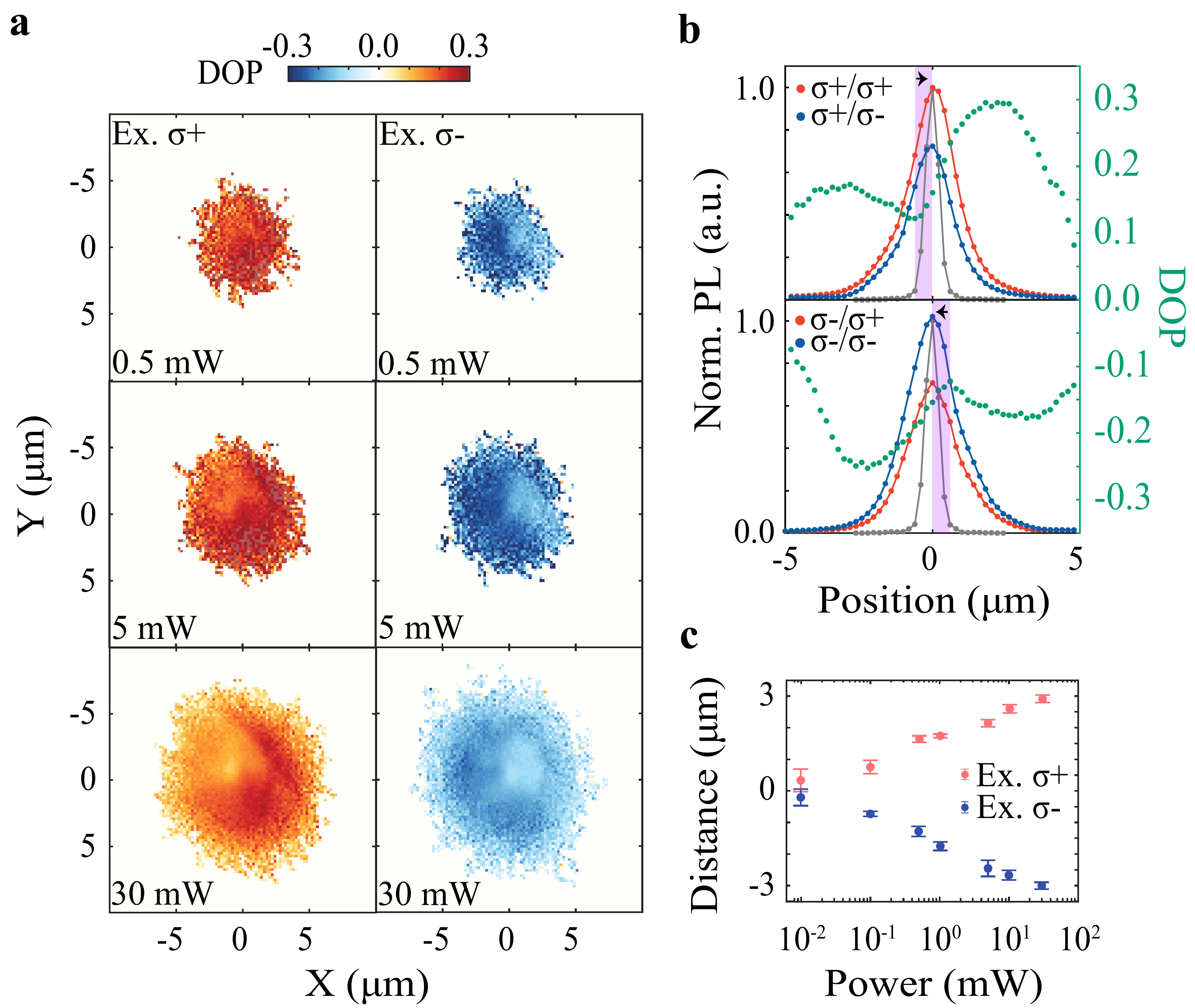}
\caption{\textbf{Drift of valley polarization.} \textbf{a,} Real-space maps of the valley polarization at excitation powers of 0.5, 5 and $30~\mathrm{mW}$ under $\sigma^{+}$ (left) and $\sigma^{-}$ (right) excitation. \textbf{b,} Polarization-resolved horizontal spatial profiles at $30~\mathrm{mW}$ excitation power for  $\sigma^{+}$ (red) and $\sigma^{-}$ (blue) detection channels. The green curve represents the resulting valley polarization profile along the horizontal direction, while the grey curve denotes the spatial profile of the excitation laser. Arrows indicate the directions of valley-polarized deflection. Purple shaded regions highlight the displacement between the excitation centre and the position of the extremum of the valley polarization, which defines the transverse valley Hall drift in real space. \textbf{c,} Drift distance of the valley-polarized signal as a function of excitation power, revealing a spatial separation of opposite valleys that follows an approximately linear dependence on a logarithmic power scale.}\label{fig3}
\end{figure}

To further investigate the anomalous optical valley Hall effect, we measured the real-space distribution of valley polarization under circularly polarized excitation. As shown in Fig.~\ref{fig3}a, the valley-polarized polaritons form a distinct ring-like spatial pattern, whose diameter increases with pump power, indicating an outward drift of the valley-polarized population. By fitting the spatial profiles with a double-Gaussian function, we extract the drift distance as a function of excitation power (Fig.~\ref{fig3}c). The drift distance grows nearly linearly with the logarithm of pump power, reaching approximately $3~\mathrm{\upmu m}$ at the highest power investigated (full dependence is shown in Extended Data Fig.~\ref{extended3}). While the origin of this increase remains to be explored -- potentially arising from local heating or enhanced many-body interactions -- this behavior highlights the tunability of valley-polariton transport. To our knowledge, the observed ring-like drift has not been reported in polaritonic systems; analogous phenomena have only been observed in interlayer excitons, where nanosecond lifetimes enable long-range diffusion~\cite{rivera2016valley}. Our results therefore constitute the first experimental demonstration of drift in valley-polarized polaritons, enabled by their ultralight effective mass and strong photonic component.

In addition to the overall outward drift, the spatial distribution of valley polarization exhibits a pronounced asymmetry. Under $\sigma^{+}$ excitation, the right side of the ring shows enhanced polarization, whereas under $\sigma^{-}$ excitation, the left side is more strongly polarized. A representative degree of polarization (DOP) profile at $30~\mathrm{mW}$ pump power is shown in Fig.~\ref{fig3}b, revealing both the ring-like drift and the lateral asymmetry across the excitation spot. Notably, the DOP minimum is displaced by approximately $0.5~\mathrm{\upmu m}$ from the pump center. Analogous to the anomalous Hall effect, we interpret this result as arising from a valley-dependent, strain-induced effective magnetic field that drives polariton transport. Our findings thus consistently demonstrate the directional steering of polariton valley currents.

\begin{figure}[t!]
\centering
\includegraphics[width=\textwidth]{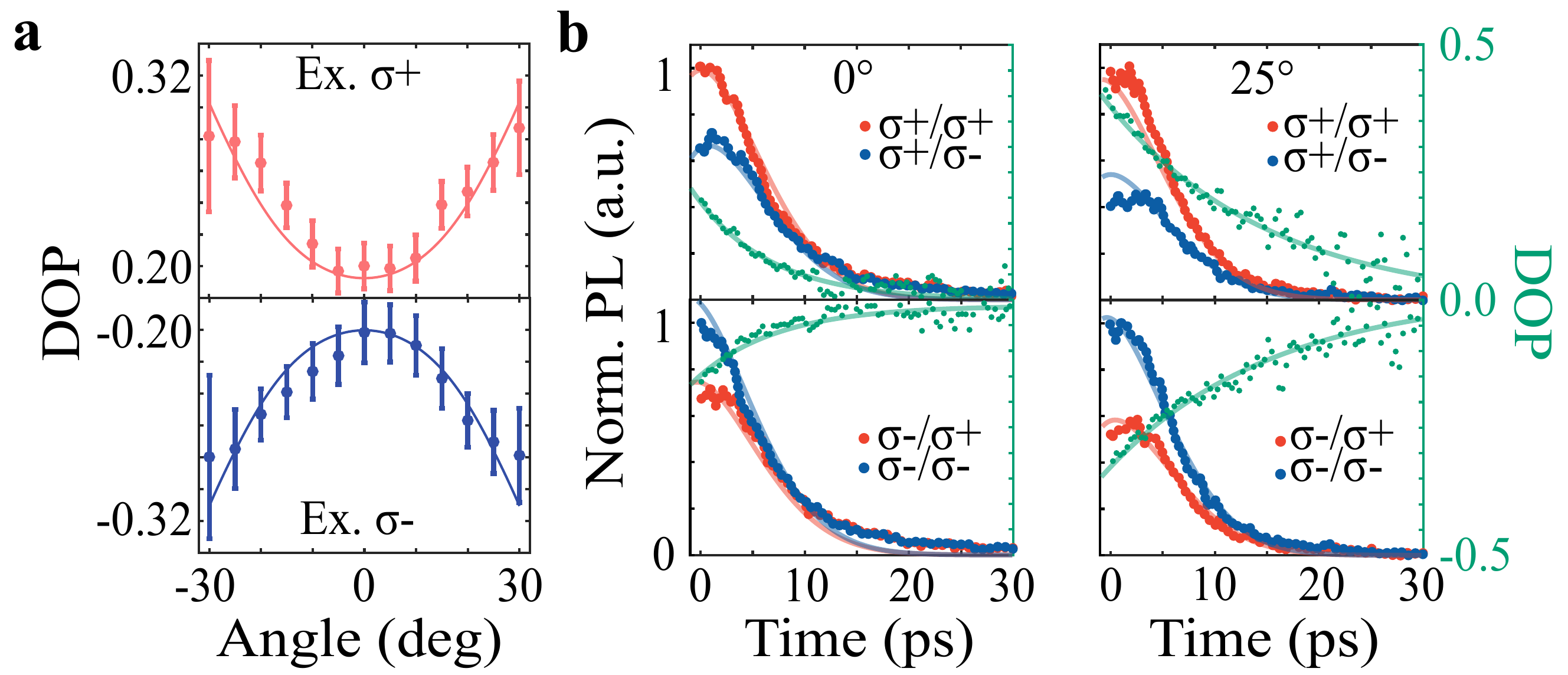}
\caption{\textbf{Angle-resolved valley polarization dynamics of polaritons.} \textbf{a,} Angle-resolved degree of polarization (DOP) under circularly polarized excitation, showing enhanced helicity at larger emission angles for both excitation helicities. \textbf{b,} Time-resolved PL at emission angles of $0^{\circ}$ and $25^{\circ}$ under $\sigma^{+}$ (top) and $\sigma^{-}$ (bottom) excitation. Red and blue dotted curves indicate PL detected in the $\sigma^{+}$ and $\sigma^{-}$ channels, respectively, while solid curves represent theoretical fits. Green dotted curves show the temporal decay of the valley polarization, with single-exponential fits indicated by solid lines. Reversal of the DOP upon switching the pump helicity is observed at both angles. The extracted lifetimes are $7.26~\mathrm{ps}$, $15.51~\mathrm{ps}$ for $\sigma^{+}$ excitation at $0^{\circ}$ and $25^{\circ}$, respectively, and $7.73~\mathrm{ps}$ and $15.84~\mathrm{ps}$ for $\sigma^{-}$ excitation at $0^{\circ}$ and $25^{\circ}$, respectively.} \label{fig4}
\end{figure}

Beyond extrinisic symmetry breaking, the manifestation of the anomalous optical valley Hall effect is critically influenced by the valley lifetime, which limits the spatial propagation of polarization. Although valley lifetimes have been extensively investigated in excitonic systems~\cite{zhu2014exciton,singh2016long,rivera2016valley}, their behavior in polaritonic platforms has remained largely unexplored. In the final section, we therefore examine the angle-resolved valley lifetime dynamics of exciton–polaritons. Owing to the strong correspondence between emission angle and polariton dispersion, this approach provides direct insight into how the photonic and excitonic fractions -- encoded in the Hopfield coefficients -- govern valley polarization dynamics. Angle-resolved measurements of the degree of polarization under $\sigma^{+}$ and $\sigma^{-}$ excitation reveal a pronounced enhancement of valley polarization at larger emission angles (Fig.~\ref{fig4}a). To quantify this effect, we performed time-resolved photoluminescence measurements at emission angles of $0^{\circ}$ and $25^{\circ}$. As shown in Fig.~\ref{fig4}b, the decay dynamics of the co- and cross-polarized PL are well described by single-exponential fits, from which the valley polarization lifetimes are extracted. Notably, the lifetime at $25^{\circ}$ is nearly twice that at $0^{\circ}$, reaching $15.51~\mathrm{ps}$ under $\sigma^{+}$ excitation, while the instrument response time remains short ($4~\mathrm{ps}$), well below the measured lifetimes. Comparable behavior is observed under $\sigma^{-}$ excitation, demonstrating the robustness of the effect. Additional measurements at $-25^{\circ}$ (Extended Data Fig.~\ref{extended2}) exhibit similar trends, further confirming the angle dependence of the valley lifetime.

To interpret the data in Fig.~\ref{fig4}, we employ the theoretical framework developed in Ref.~\cite{dufferwiel2017valley}. Within this approach, optical excitation first creates an excitonic reservoir, which subsequently relaxes into polariton states; the experimentally detected signal arises from the radiative decay of these polaritons. Guided by our observations, we model the system using a single effective polariton branch. This approximation yields a set of coupled rate equations that can be solved analytically or numerically. Most parameters are adopted from Ref.~\cite{dufferwiel2017valley}, with the exception of the spin relaxation times, which are adjusted to reflect the faster photonic spin relaxation ($1.3~\mathrm{ps}$) and slower excitonic reservoir valley relaxation ($5~\mathrm{ps}$) observed in our system. Our time scales are consistent with Ref.~\cite{sun2017optical}. A full description of the model and fitting procedure is provided in the Supplementary Materials.

The strong agreement between theory and experiment allows us to draw a number of conclusions regarding the angular dependence observed in Fig.~\ref{fig4}a. The enhanced degree of polarization at larger emission angles originates from the evolving balance between the excitonic and photonic components of the polaritons. This balance is quantified by the Hopfield coefficients, which enter the model in two essential ways: they control the relaxation rate from the excitonic reservoir into polariton states and determine the polariton lifetime and spin relaxation dynamics. Both contributions are crucial for accurately describing the experimental observations. At larger in-plane momenta, polaritons acquire a stronger excitonic character, leading to more efficient population from the excitonic reservoir while simultaneously suppressing depolarization. As a result, the degree of polarization increases with emission angle. More generally, the model predicts a monotonic increase of the coherence time with angle, in excellent agreement with the time-resolved measurements shown in Fig.~\ref{fig4}b. Notably, polaritons emitted at larger angles exhibit a long valley lifetime of approximately $15~\mathrm{ps}$ together with higher group velocities. These long valley lifetimes are one of the key factors enabling the observation of the anomalous optical valley Hall effect. 

{

In summary, we have experimentally demonstrated an anomalous optical valley Hall effect in a monolayer WS${}_2$ exciton-polariton microcavity. Polarization- and time-resolved real-space imaging reveals a symmetry-breaking spatial separation of valley-polarized polaritons, accompanied by a Hall drift velocity approaching $10^{5}~\mathrm{m/s}$, underscoring the exceptional dynamical response enabled by hybrid light–matter coupling. This high-speed valley Hall transport originates from the ultralight effective mass of polaritons, which facilitates efficient valley encoding and robust propagation over extended length scales. By establishing a direct and scalable analogue of the conventional Hall effect in a polaritonic platform, our results position exciton-polaritons as a powerful framework for ultrafast valley manipulation. More broadly, they open opportunities for reconfigurable valleytronic and topological photonic functionalities based on engineered light-matter interactions.
}

\section{Methods}

\textbf{Fabrication}. The bottom distributed Bragg reflector (DBR), comprising 30 alternating SiO${}_2$/TiO${}_2$ layers on a silicon substrate, was fabricated by E-beam Evaporator (SHINCRON, MIC-1350TBN). A $105~\mathrm{nm}$ SiO${}_2$ spacer was then deposited on the DBR using plasma-enhanced chemical vapor deposition (PECVD PlasmaPro 100). A monolayer WS${}_2$, grown by chemical vapor deposition (CVD), was transferred onto the surface via a water-intercalation technique (see Ref. ~\cite{yu2024high} for details of growth and transfer). Subsequently, a $70~\mathrm{nm}$ polymethyl methacrylate (PMMA) layer was spin-coated, and a $50~\mathrm{nm}$ silver film was deposited by thermal evaporation to serve as the top mirror.

\textbf{Optical Measurements}. All measurements were carried out at $5~\mathrm{K}$ in a closed-cycle cryostat (CryoAdvance 50 base, Montana Instruments). Angle-resolved reflectance spectra were obtained using a home-built Fourier imaging setup with a broadband white-light source, where a pinhole aperture defined a measurement spot of $\sim3~\mathrm{\upmu m}$. An achromatic objective (NA = 0.41, 40×) was used for both illumination and collection. The reflected signal was directed into a spectrometer (HRS500, Princeton Instruments) through a narrow entrance slit and detected by a two-dimensional CCD camera (Blaze, Princeton Instruments). For polarization-resolved PL, circularly polarized excitation was generated by a linear polarizer and a quarter-wave plate in the excitation path, while the emitted PL was analyzed using a quarter-wave plate and a linear polarizer before the spectrometer. Excitation was provided by a $532~\mathrm{nm}$ femtosecond pulsed laser (Chameleon, Coherent; $150~\mathrm{fs}$, $80~\mathrm{MHz}$). Time-resolved PL was measured with a streak camera (C16910, Hamamatsu).

\begin{backmatter}
\section{Data availability}
    All data in the manuscript are available from the corresponding author upon reasonable request.
    
\section{Acknowledgment}
    The work is supported by the National Key Research and Development
    Program of China (2021YFA1200803); the Innovation Program for Quantum Science and Technology (2023ZD0300300), the National Natural Science Foundation of China (12174111), Shanghai Pilot Program for Basic Research (TQ20240204), and the Science and Technology Commission of Shanghai Municipality (23JC1402000). 

\section{Author contributions}
    Z.S. and J.W. initiated the project. XZ.C., LX.Y., and SR.S. fabricated the samples. XZ.C., YJ.G., and Z.S. performed the experiments and carried out the simulations. G.A. and V.A. performed the numerical simulation. XZ.C., YJ.G., G.A., V.A. and Z.S. analyzed the data; All the authors contributed to the discussion of the results and wrote the manuscript.
    
\section{Competing interests}
    The authors declare no competing interests.
    
\end{backmatter}

\bibliography{sn-bibliography}
\newpage

\setcounter{figure}{0}
\renewcommand{\figurename}{Extended Data Fig.}
\makeatletter
\renewcommand{\fnum@figure}{\bfseries\figurename\thefigure}
\makeatother

\begin{figure}[H]
\centering
\includegraphics[width=\textwidth]{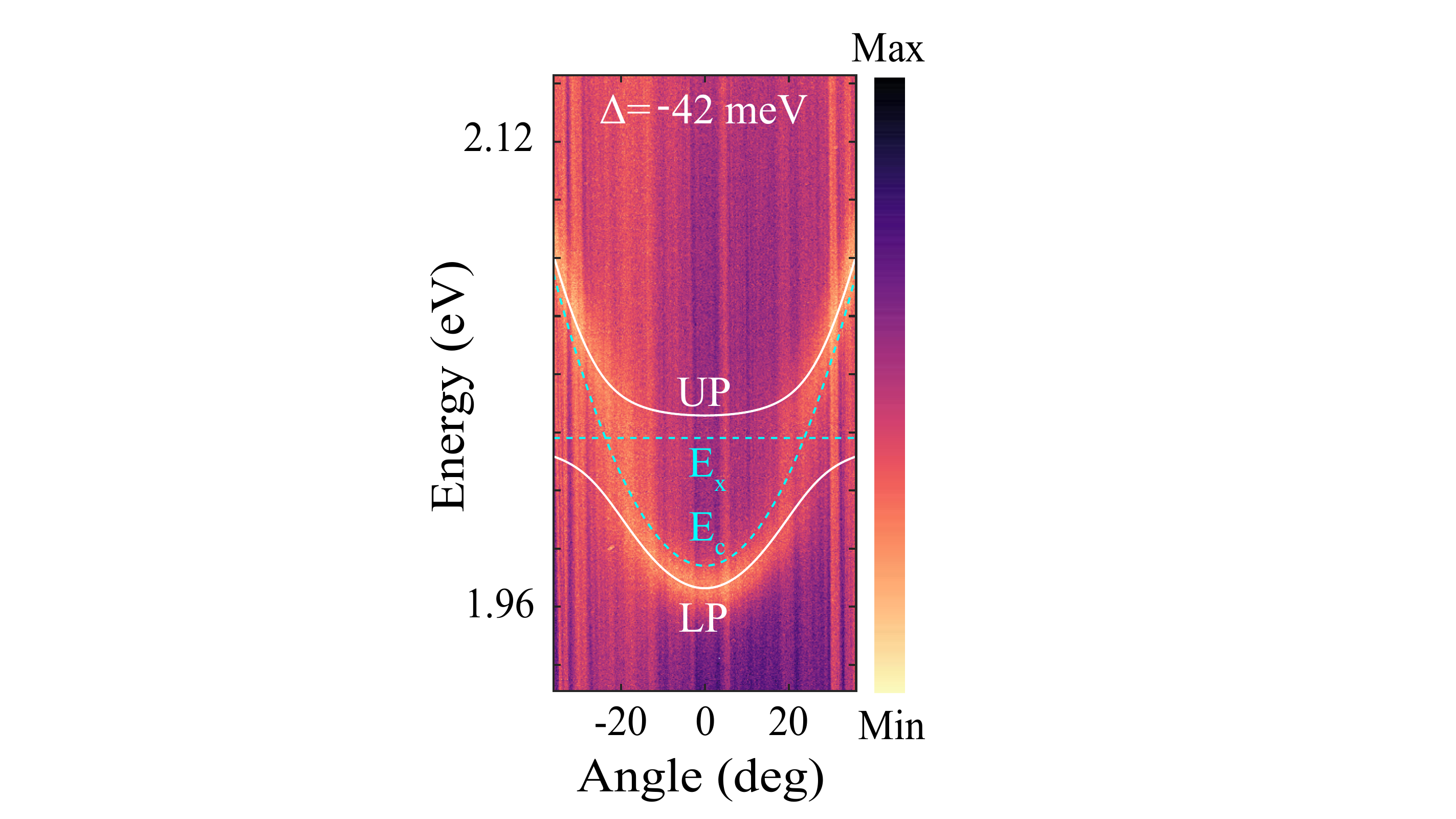}
\caption{\textbf{Angle-resolved reflectance spectra of the sample measured at room temperature.} A distinct anticrossing behavior is observed, confirming the strong coupling regime in the system. As the temperature increases to room temperature, the exciton energy exhibits a redshift, resulting in a detuning value of $-42 ~\mathrm{meV}$. The solid white curves denote the upper (UP) and lower polariton (LP) branches obtained from a coupled-oscillator fit, while the blue dashed lines mark the exciton resonance, $E_X$, and the bare cavity photon dispersion, $E_c$.}\label{extended1}
\end{figure}

\begin{figure}[H]
\centering
\includegraphics[width=0.9\textwidth]{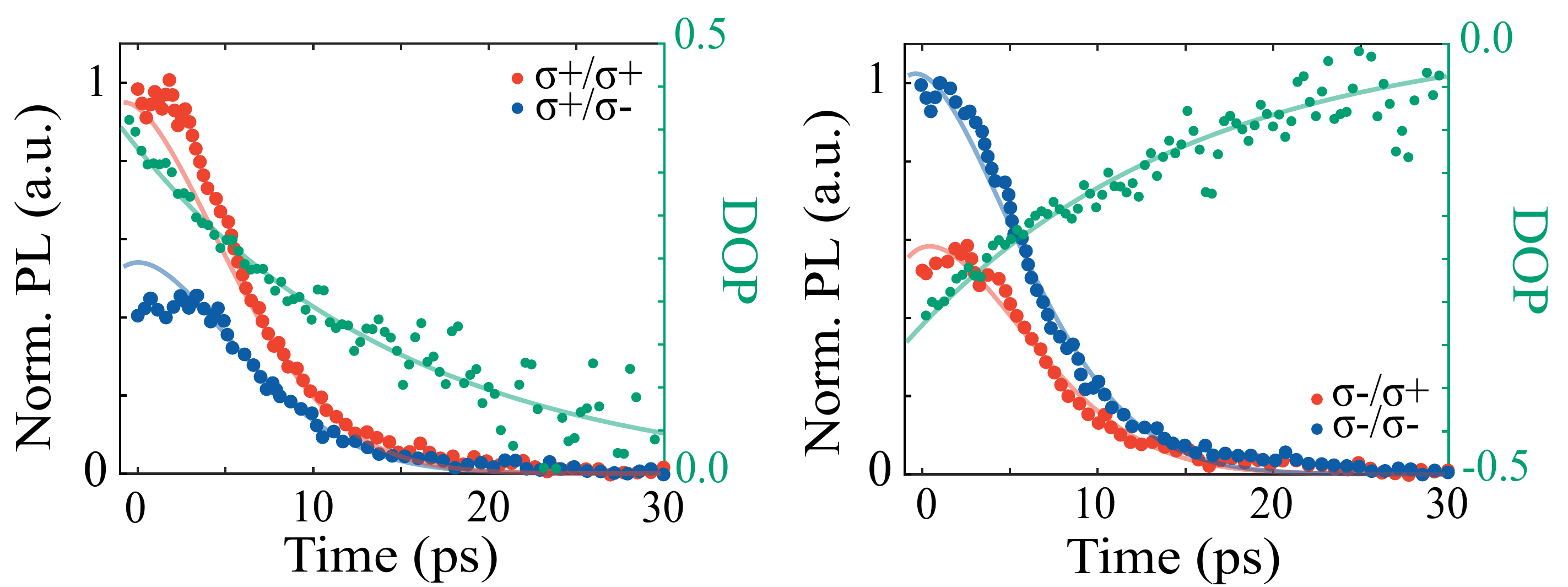}
\caption{\textbf{Time-resolved PL at -$25^\circ$ under circular excitation.} Red and blue dotted curves correspond to PL collected in the $\sigma^+$ and $\sigma^-$ channels, respectively, while solid curves represent theoretical fits. Green dotted curves show the temporal decay of valley polarization with single-exponential fits (solid lines). The extracted lifetimes are 15.75 and $15.59~\mathrm{ps}$.}\label{extended2}
\end{figure}

\begin{figure}[H]
\centering
\includegraphics[width=\textwidth]{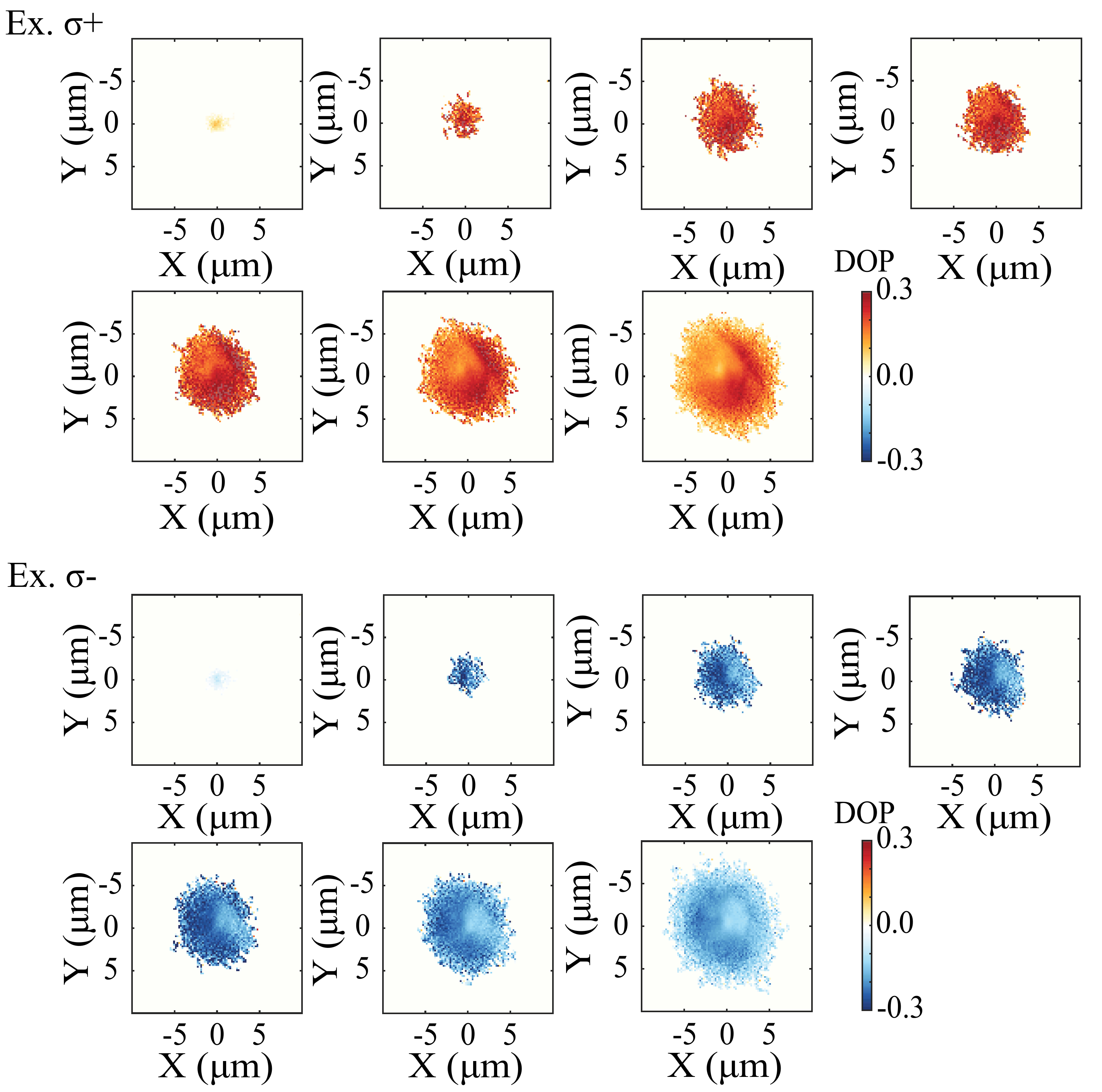}
\caption{\textbf{The real-space valley polarization distribution under $\sigma^+$ and $\sigma^-$ optical excitation.} From left to right and top to bottom, the corresponding excitation powers are 0.01, 0.1, 0.5, 1, 5, 10, and $30~\mathrm{mW}$. The images reveal a ring-like spatial pattern whose diameter increases with rising excitation power.}\label{extended3}
\end{figure}

\begin{figure}[H]
\centering
\includegraphics[width=\textwidth]{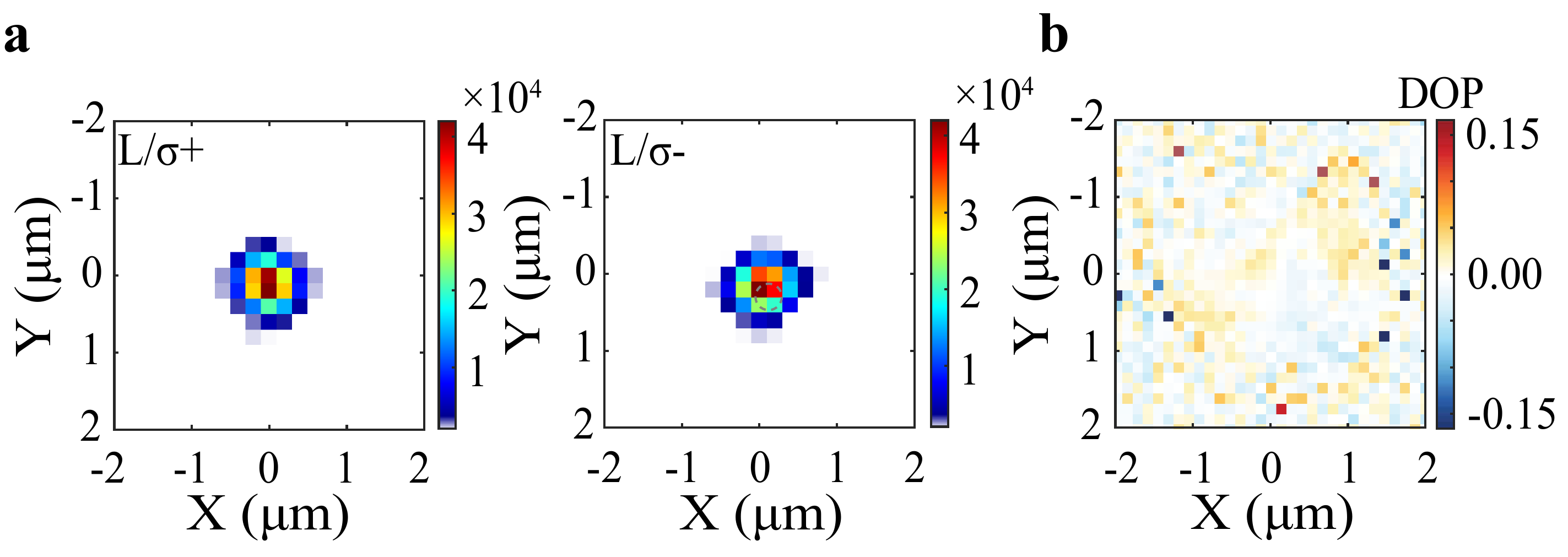}
\caption{\textbf{The real-space DOP distribution of the incident laser.} \textbf{a,} The real-space $\sigma+$ and $\sigma-$ components distribution of reflected linearly polarized laser, acquired under the same collection configuration as used in the PL measurements. \textbf{b,} The real-space distribution of DOP calculated from \textbf{a}. The absence of circular polarization separation confirms that the anomalous optical valley Hall effect observed in the experiments is not an artifact introduced by the optical path or instrumental configuration.}\label{extended4}
\end{figure}

\begin{figure}[H]
\centering
\includegraphics[width=\textwidth]{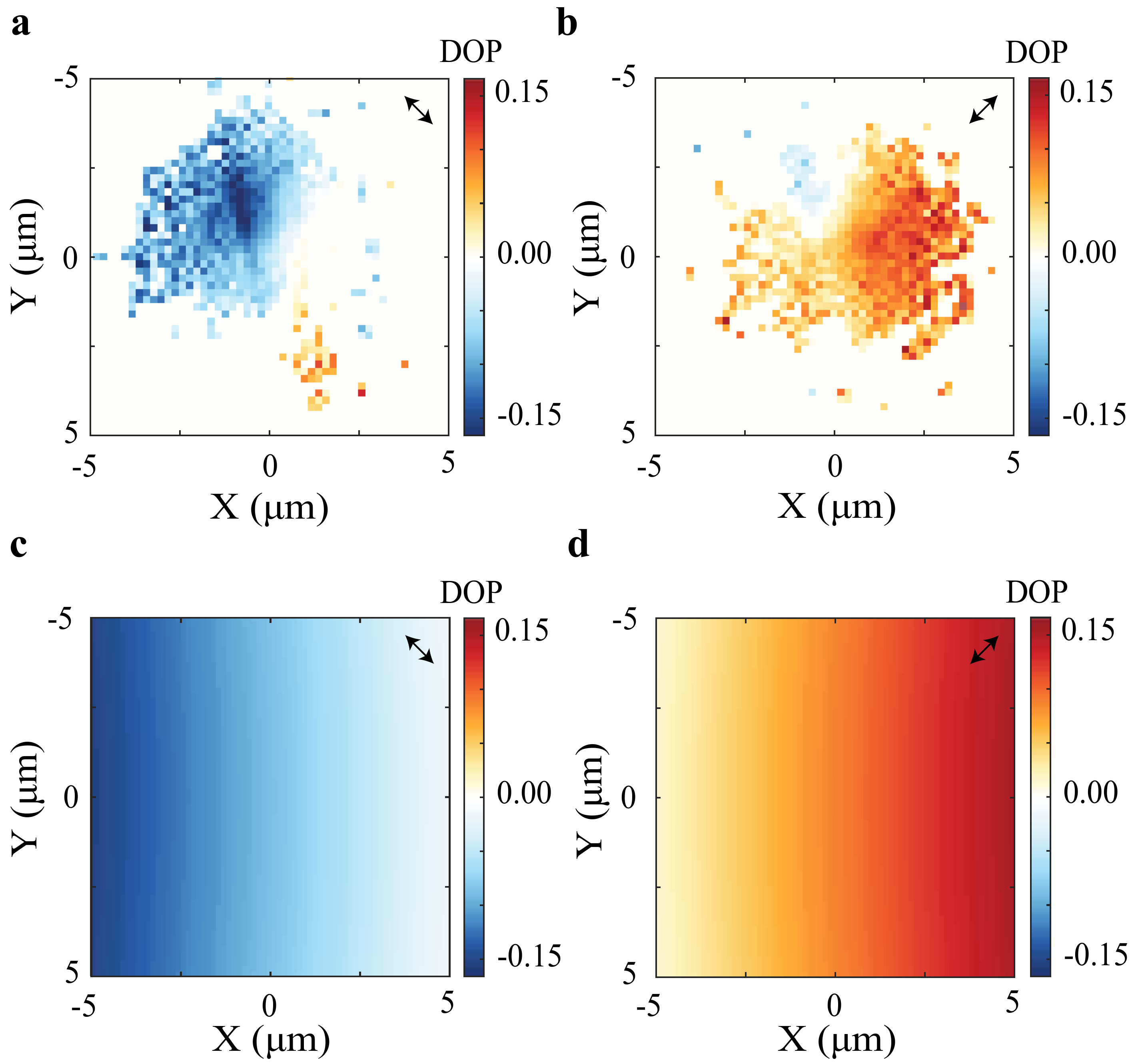}
\caption{\textbf{Real-space valley polarization distribution under linear excitation of $-45^\circ$ and $45^\circ$.} \textbf{a,} \textbf{b,} The experimental real-space valley polarization distribution under linear excitation of -$45^\circ$ and $45^\circ$. \textbf{c,} \textbf{d,} The simulated real-space valley polarization distribution under linear excitation of $-45^\circ$ and $45^\circ$. The arrows represent the angles of linear excitation.}\label{extended5}
\end{figure}

\newpage

\setcounter{figure}{0}
\renewcommand{\figurename}{Fig.S}
\makeatletter
\renewcommand{\fnum@figure}{\bfseries\figurename\thefigure}
\makeatother

\textbf{Theoretical model for dynamics of the anomalous optical valley Hall effect}

\vspace{1em}

To interpret the anomalous Hall effect, we rely on the strain-induced magnetic field. According to this hypothesis, the Hamiltonian that describes time dynamics of polariton is written as
\begin{equation}
H=\frac{\hbar^2k^2}{2m_p}+\frac{\hbar}{2}\sum_{i=1}^3 \sigma_i \cdot\Omega_i, 
\end{equation}
where the components of the magnetic field are~\cite{glazov2022exciton}
\begin{equation}
\Omega_x=b(u_{xx}-u_{yy}), \qquad \Omega_y=2bu_{xy}, \qquad \Omega_z=c[(u_{xx}-u_{yy})k_x-2u_{xy}k_y].
\end{equation}
For the sake of simplicity, we have disregarded the `intrinsic' magnetic fields due to the longitudinal-transversal splitting of the excitonic states and TL-TM splitting of the cavity modes. This approximation is justified as our experimental data suggests that the strain leads to the strongest effective magnetic field. 
Furthermore, this approximation allows us to reproduce experimental data within a minimal model. There are no known estimates for the parameters $b$ and $c$ in a polaritonic systems. Therefore,
we used $\hbar b=1~\mathrm{meV}$, $\hbar c=0.4~\mathrm{meV}$$\cdot\upmu$m 
consistent with the values for the excitonic systems in TMDs~\cite{durnev2018excitons,glazov2022exciton,yagodkin2024excitons}. Using these values together with reasonable strains $(u_{xx}-u_{yy})=0.2$\% and $u_{xy}=0.1$\%, we are able to describe the data qualitatively. 

To incorporate the finite lifetimes of polariton quasiparticles, two closely related approaches can be considered. The first employs the Hamiltonian to derive a differential equation for the evolution of the density matrix, explicitly including the decay rate (see, e.g.,~\cite{kavokin2004quantum}). The second approach involves solving the Schr{\"o}dinger equation, with the decay rate introduced later when computing observables from the solution~\cite{leyder2007observation}. As we checked that both approaches lead to similar results in momentum space, we use the second approach in our calculations, as it simplifies real-space calculations.  
Namely, we first solve the Schr{\"o}dinger equation in momentum space
\begin{equation}
i\hbar\frac{\partial}{\partial t}\begin{pmatrix}
a_+\\
a_-
\end{pmatrix}
=H \begin{pmatrix}
a_+ \\
a_-
\end{pmatrix},
\end{equation}
where $a_+$ and $a_-$ define the quasi-spin populations. As the initial condition, we use $a_{+}=1/\sqrt{2}$ and $a_-=e^{i\phi}/\sqrt{2}$, where the parameter $\phi$ determines the orientation of the linear polarization of light with respect to the direction of the strain. Then, we calculate the corresponding time dynamics in real space: 
\begin{equation}
\psi_{\pm}(x,y,t)=\int a_{\pm}(k_x,k_y,t)e^{i k_x x + i k_y y}e^{-\frac{k_x^2+k_y^2}{k_0^2}}\mathrm{d}k_x\mathrm{d}k_y,
\end{equation}
where $k_0$ defines the initial distribution of polaritons in momentum space. To estimate $k_0$, we note that the Hall velocity within our theory can be approximated by $\hbar k_0/m_p$, because the parameter $k_0$ is the natural parameter of the speed of information propagation. The experimental data suggests that $k_0$ should be of the order of $0.1~\mathrm{\upmu m^{-1}}$. We use this value in our calculations. Note that this value is smaller than the scale given by the laser, $\sim 0.5~\mathrm{\upmu m^{-1}}$. This is natural, since equilibration drives the polariton population toward smaller momenta. Overall, we observe that variations in $k_0$ do not quantitatively affect our conclusions. 

Once the functions $\psi_{\pm}(x,y)$ are calculated, we compute the PL signal as 
\begin{equation}
I(x,y)=\delta_I \frac{\int_0^\infty \mathrm{d}t e^{-t/\tau_p}(|\psi_{+}(x,y,t)|^2-|\psi_{-}(x,y,t)|^2)}{\int_0^\infty \mathrm{d}t e^{-t/\tau_p}(|\psi_{+}(x,y,t)|^2+|\psi_{-}(x,y,t)|^2)},
\end{equation}
where the exponent $ e^{-t/\tau_p}$ accounts for a finite lifetime 
of a polariton. The parameter $\delta_I$ is introduced to better fit the data. We observed that it is around 2-3 for the considered cases. 

\begin{figure}[H]
\centering
\includegraphics[width=\textwidth]{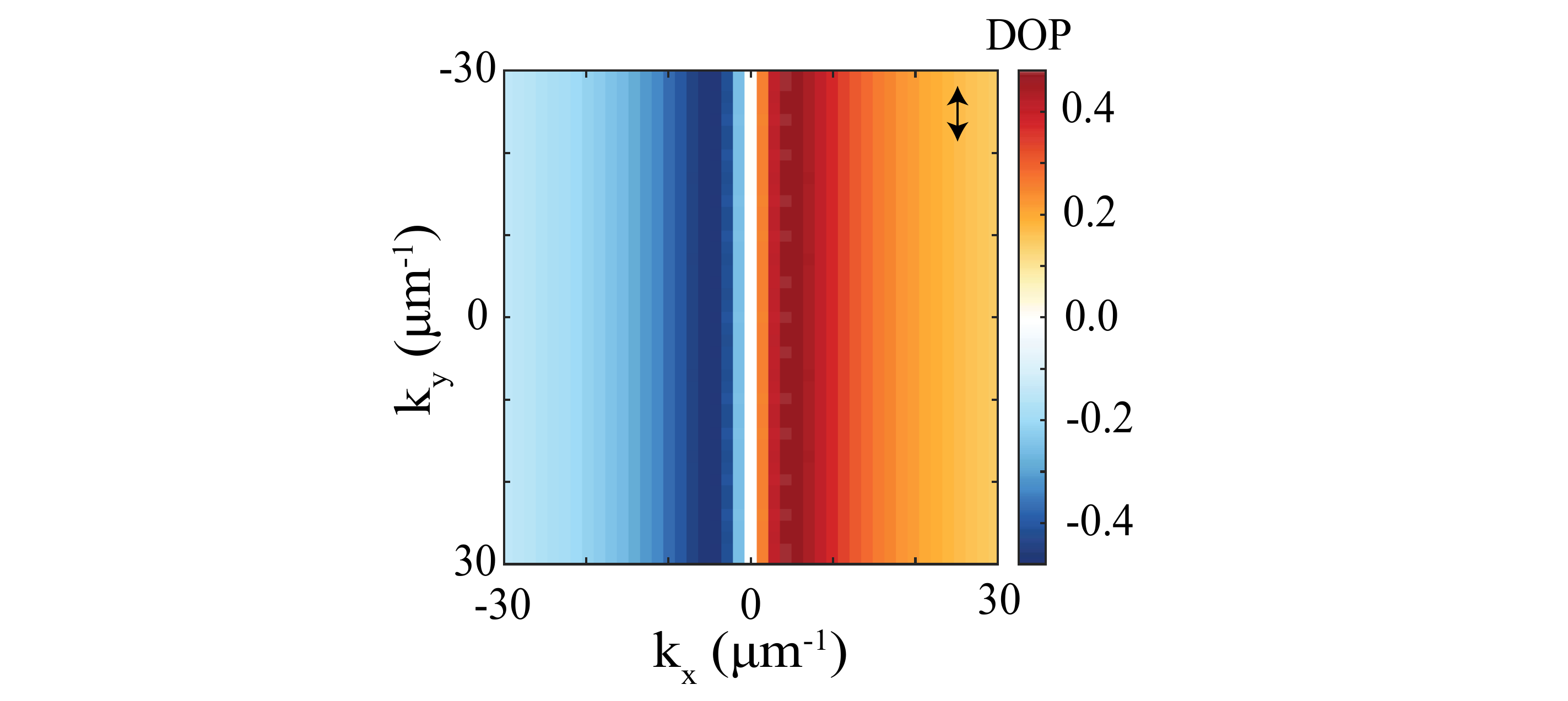}
\caption{\textbf{The simulated valley polarization distribution in k space.} Polaritons at $-\mathrm{k}_x$ and $\mathrm{k}_x$ exhibit negative and positive valley polarization, respectively.}\label{figS3}
\end{figure}

\textbf{Theoretical model for angle-resolved valley polarization dynamics}

\vspace{1em}

Motivated by the theories used in Refs.~\cite{dufferwiel2017valley,sun2017optical}, we solve the rate equations 
\begin{align}
\frac{\mathrm{d}n_{X+}}{\mathrm{d}t}=P_{+}-\frac{n_{X+}}{\tau_X}-n_{X+}W(\theta)-\frac{n_{X+}-n_{X-}}{\tau_{X\pm}},\nonumber\\
\frac{\mathrm{d}n_{X-}}{\mathrm{d}t}=P_{-}-\frac{n_{X-}}{\tau_X}-n_{X-}W(\theta)+\frac{n_{X+}-n_{X-}}{\tau_{X\pm}},\nonumber\\
\frac{\mathrm{d}n_{LP+}}{\mathrm{d}t}=W(\theta)n_{X+}-\frac{n_{LP+}}{\tau_{LP}(\theta)}-\frac{n_{LP+}-n_{LP-}}{\tau_{LP\pm}(\theta)},\nonumber\\
\frac{\mathrm{d}n_{LP-}}{\mathrm{d}t}=W(\theta)n_{X-}-\frac{n_{LP-}}{\tau_{LP}(\theta)}+\frac{n_{LP+}-n_{LP-}}{\tau_{LP\pm}(\theta)},
\label{eq:rate_SM}
\end{align}
where $n_{X\pm}$ describes the population of the excitonic bath, and $n_{LP\pm}$ describes the population of the lowest polaritonic branch. To minimize the number of fitting parameters, we disregard other branches (e.g., trion, upper polariton branches) in our analysis. 
As our experiment operates with a 150 fs laser, we assume that the source terms, $P_{\pm}$, are vanishing. Instead, we solve the system of equations with relevant initial conditions. For example, for a $\sigma_{+}$-excitation we use $n_{X+}=1$ and $n_{X-}=n_{LP+}=n_{LP-}=0$ at $t=0$.

\begin{figure}[H]
\centering
\includegraphics[width=\textwidth]{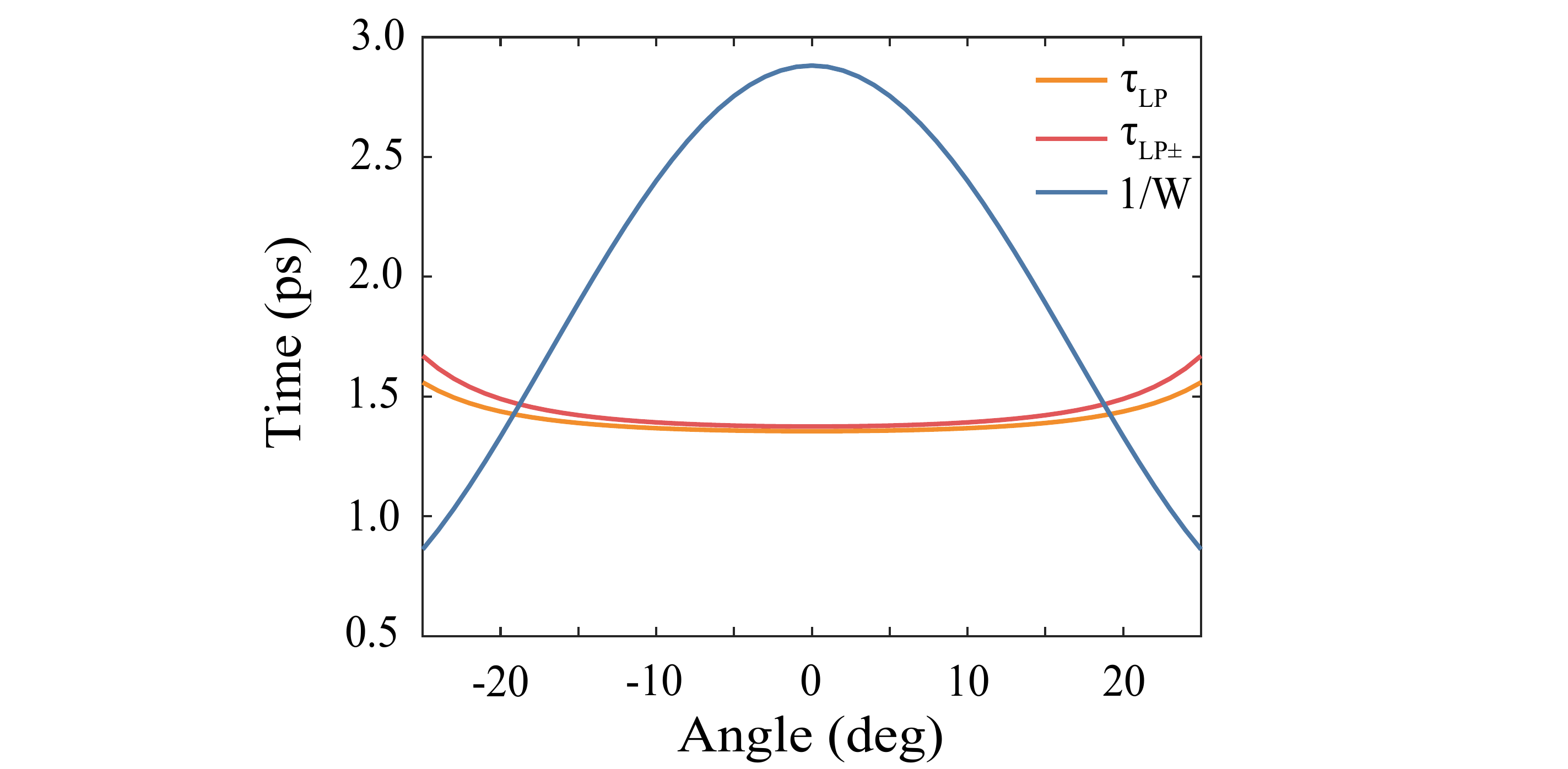}
\caption{\textbf{\bm{$\tau_{LP}(\theta)$}, \bm{$\tau_{LP\pm}(\theta)$}, \bm{$1/W(\theta)$} as a function of angle.} The quantities $\tau_{LP}(\theta)$ and $\tau_{LP\pm}(\theta)$ increase with the angle, whereas $1/W(\theta)$ decreases as the angle increases.}\label{figS1}
\end{figure}

The timescales in the system of equations are also motivated by Refs.~\cite{dufferwiel2017valley,sun2017optical}. Namely, we use $\tau_X=5~\mathrm{ps}$, and
\begin{align}
\frac{1}{\tau_{LP}(\theta)}=\frac{X(\theta)}{\tau_X}+\frac{P(\theta)}{\tau_p},\qquad
\frac{1}{\tau_{LP\pm}(\theta)}=\alpha\frac{X(\theta)}{\tau_{X\pm}}+\frac{P(\theta)}{\tau_{p\pm}},
\end{align}
where $\alpha$ serves to account for the attenuation of polariton spin relaxation arising from its excitonic component, relative to that observed in a bare flake exciton. We use $\alpha=0.05$~\cite{dufferwiel2017valley}. However, we observed that our results are largely insensitive to its exact value, in particular, because we used a large value of $\tau_{X\pm}$ in comparison to Ref.~\cite{dufferwiel2017valley} to account for the increased lifetime of the polariton. Following, Ref.~\cite{sun2017optical}, we use $\tau_{X\pm}=5~\mathrm{ps}$. For the photonic component, we used $\tau_{p}=\tau_{p\pm}=1.3~\mathrm{ps}$. The Hopfield coefficients 
$X(\theta)$ and $P(\theta)$ are
\begin{equation}
X(\theta)=\frac{1}{2}+\frac{1}{2}\frac{\Delta_{\theta}}{\sqrt{\Delta_{\theta}^2+4g^2}}, \qquad P(\theta)=1-X(\theta),
\end{equation}
where the angle-dependent detuning is $\Delta_{\theta}=\Delta+0.288 \text{eV}\sin(\theta\pi/180)^2$ with $\Delta=-0.078~\mathrm{eV}$, and $\theta$ measured in grad; $g=0.02~\mathrm{eV}$. These parameters are extracted from the angle-resolved reflectance spectra using a coupled-oscillator model, as illustrated in the main text. The final ingredient for the system of equations~(\ref{eq:rate_SM}) is the scattering rate $W(\theta)$. Reference~\cite{dufferwiel2017valley} provides a simple expression $W(\theta)=W_0 X(\theta)P(\theta)$, which we utilize in our work as well, with $1/W_0=0.15~\mathrm{ps}$. The relevant timescales for Eq.~(\ref{eq:rate_SM}) are illustrated in Fig.~S\ref{figS1}. 

The system of equations~(\ref{eq:rate_SM}) describes our data in Fig. 4a of the main text. To describe the time-resolved PL, we need to include additional information, such as the instrument response time. To this end, we calculate a convolution
\begin{equation}
PL(t)\sim \int_{-\infty}^t n_{LP\pm}(\tau)e^{-\frac{(\tau-t)^2}{2 \tau_r^2}}\mathrm{d}\tau.
\end{equation}
We find that the best fit to our data is obtained with $\tau_r \approx 7~\mathrm{ps}$, close to the expected response time of 4 ps.

\begin{figure}[H]
\centering
\includegraphics[width=\textwidth]{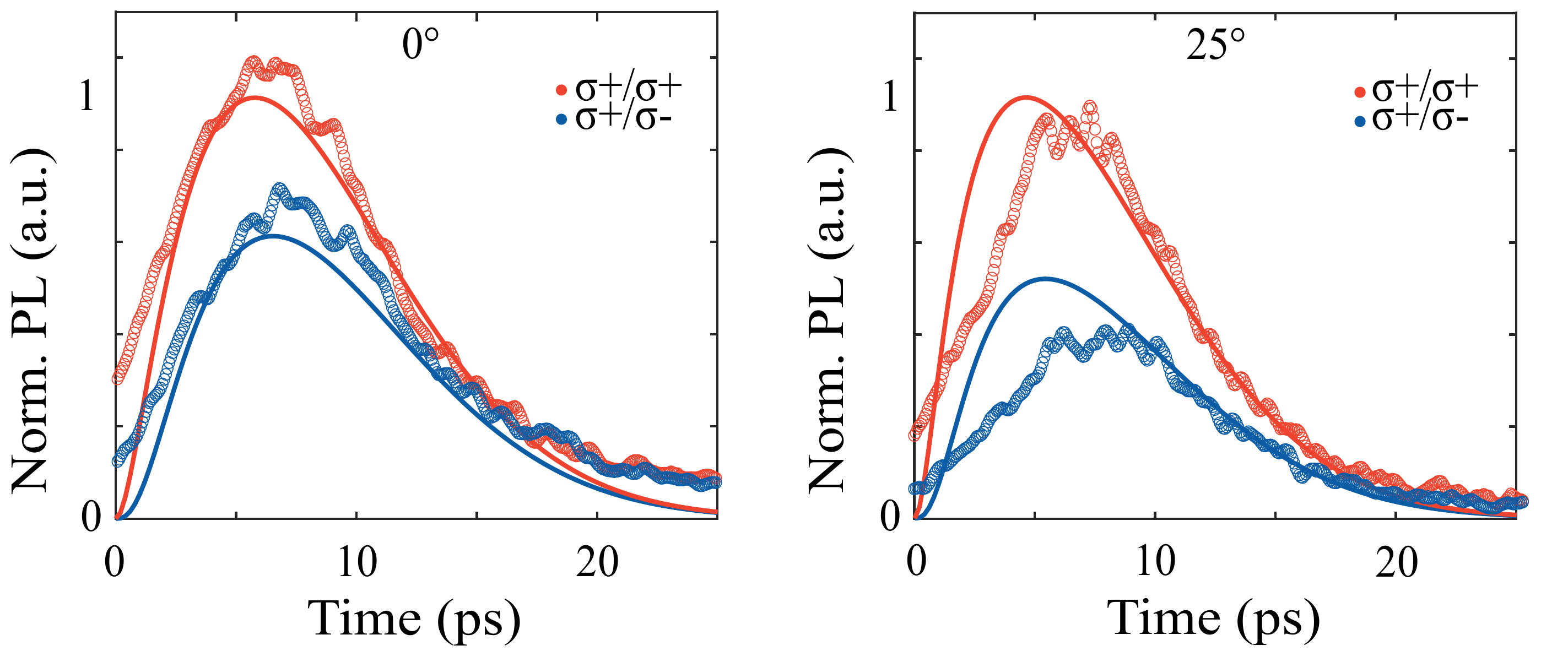}
\caption{\textbf{Theoretical simulation and experimental data of time-resolved PL intensity.} The theoretical simulation (solid lines) and experimental data (circles) of $\sigma^+$ and $\sigma^-$ PL intensity as a function of time under $\sigma^+$ excitation. Left: $0^\circ$. Right: $25^\circ$. }\label{figS2}
\end{figure}

\vspace{1em}

\end{document}